\documentclass[aps,prd,superscriptaddress,showpacs,preprintnumbers]{revtex4}
\usepackage{graphicx,color}

\newcommand{\be}{\begin{equation}}
\newcommand{\ee}{\end{equation}}
\newcommand{\bea}{\begin{eqnarray}}
\newcommand{\eea}{\end{eqnarray}}

\newcommand{\bfk}{\mbox{\boldmath $k$}}
\def\bkt{\bfk_\perp}

\newcommand{\bfp}{\mbox{\boldmath $p$}}

\def\bpp{\bfp_\perp}

\newcommand{\bfP}{\mbox{\boldmath $P$}}
\newcommand{\bfS}{\mbox{\boldmath $S$}}
\newcommand{\bfs}{\mbox{\boldmath $s$}}

\newcommand{\pup}{p^\uparrow}

\newcommand{\qup}{q^\uparrow}

\def\lsim{\mathrel{\rlap{\lower4pt\hbox{\hskip1pt$\sim$}}\raise1pt\hbox{$<$}}}
\def\gsim{\mathrel{\rlap{\lower4pt\hbox{\hskip1pt$\sim$}}\raise1pt\hbox{$>$}}}
\def\nostrocostruttino#1\over#2{\mathrel{\mathop{\kern 0pt \rlap
{\hbox{$#1$}}} \hbox{\kern-.135em $#2$}}}

\begin{document}

\title{Single Spin Asymmetries in $\ell \, \pup \to h \, X$ processes and TMD
factorisation}

\author{M.~Anselmino}
\affiliation{Dipartimento di Fisica Teorica, Universit\`a di Torino,
             Via P.~Giuria 1, I-10125 Torino, Italy}
\affiliation{INFN, Sezione di Torino, Via P.~Giuria 1, I-10125 Torino, Italy}
\author{M.~Boglione}
\affiliation{Dipartimento di Fisica Teorica, Universit\`a di Torino,
             Via P.~Giuria 1, I-10125 Torino, Italy}
\affiliation{INFN, Sezione di Torino, Via P.~Giuria 1, I-10125 Torino, Italy}
\author{U.~D'Alesio}
\affiliation{Dipartimento di Fisica, Universit\`a di Cagliari,
             I-09042 Monserrato (CA), Italy}
\affiliation{INFN, Sezione di Cagliari,
             C.P.~170, I-09042 Monserrato (CA), Italy}
\author{S.~Melis}
\affiliation{Dipartimento di Fisica Teorica, Universit\`a di Torino,
             Via P.~Giuria 1, I-10125 Torino, Italy}
\author{F.~Murgia}
\affiliation{INFN, Sezione di Cagliari,
             C.P.~170, I-09042 Monserrato (CA), Italy}
\author{A.~Prokudin}
\affiliation{Jefferson Laboratory, 12000 Jefferson Avenue, Newport News, VA 23606, USA}
\date{\today}

\begin{abstract}
Some estimates for the transverse Single Spin Asymmetry, $A_N$, in the inclusive processes $\ell \, \pup \to h \, X$, given in a previous paper, are expanded and compared with new experimental data. The predictions are based on the Sivers distributions and the Collins fragmentation functions which fit the azimuthal asymmetries measured in Semi-Inclusive Deep Inelastic Scattering (SIDIS) processes ($\ell \, \pup \to \ell' \, h \, X$). The factorisation in terms of Transverse Momentum Dependent distribution and fragmentation functions (TMD factorisation) --  {\it i.e.} the theoretical framework in which SIDIS azimuthal asymmetries are analysed -- is assumed to hold also for the inclusive process $\ell \, p \to h \, X$ at large $P_T$. The values of $A_N$ thus obtained agree in sign and shape with the data. Some predictions are given for future experiments.

\end{abstract}


\maketitle

\section{\label{Intro} Introduction}

In a previous paper~\cite{Anselmino:2009pn} the issue of the validity of the
TMD factorisation for hard inclusive processes in which only one large scale
is detected has been investigated in a simple phenomenological approach.
We considered transverse Single Spin Asymmetries (SSAs) for the
$\ell \, \pup \to h \, X$ process, with the detection, in the lepton-proton
center of mass ({\it c.m.}) frame, of a single large $P_T$ final particle, typically
a pion. Also the case of jet production, $\ell \, \pup \to {\rm jet} + X$, was
considered. The final lepton is not necessarily observed; however, a large
value of $P_T$ implies, at leading perturbative order, large values of $Q^2$,
and the active role of a hard elementary interaction, $\ell \, q \to \ell \, q$.
Such a measurement is the exact analogue of the SSAs observed in the $p \, \pup
\to h \, X$ processes, the well known and large left-right asymmetries
$A_N$, measured over a huge energy range~\cite{Adams:1991rw,Adams:1991cs,
Adams:1991rv,Adams:1991ru,Adams:2003fx,Adler:2005in,Lee:2007zzh,
Abelev:2008af,Adamczyk:2012xd,Igo:2012,Bland:2013pkt}.
On the other hand, the process is essentially a Semi-Inclusive Deep Inelastic
Scattering (SIDIS) process, for which, at large $Q^2$ values (and small $P_T$
in the $\gamma^* - p\,$ {\it c.m.} frame), the TMD factorisation is proven to hold
~\cite{Collins:2002kn, Collins:2004nx,Ji:2004wu, Ji:2004xq, Bacchetta:2008xw}.

We computed these SSAs assuming the TMD factorisation and using the relevant
TMDs (Sivers and Collins functions) as extracted from SIDIS data. A
similar idea of computing left-right asymmetries in SIDIS processes,
although with different motivations and still demanding the
observation of the final lepton, has been discussed in
Ref.~\cite{She:2008tu}. A first simplified study of $A_N$ in $\ell
\, \pup \to h \, X$ processes was performed in
Ref.~\cite{Anselmino:1999gd}. The process was also considered in
Refs.~\cite{Koike:2002ti,Koike:2002gm} in the framework of collinear
factorisation with twist-three correlation functions, obtaining asymmetries
with a sign opposite to that of the corresponding ones in $p \, p$ processes.
Jet production in $\ell \, p \to {\rm jet} + X$ was studied in
Ref.~\cite{Kang:2011jw}, in a collinear factorisation scheme with a higher-twist
quark-gluon-quark correlator, $T_F$, which is related to the first moment of
the Sivers function~\cite{Boer:2003cm,Ma:2003ut,Zhou:2008mz,Zhou:2009jm}.

While at the time of publication of Ref.~\cite{Anselmino:2009pn}
no data were available on $A_N$ from lepton-proton inclusive processes,
very recently some experimental results have been published by
the HERMES Collaboration~\cite{Airapetian:2013bim}. New data are also
available by the JLab Hall A Collaboration~\cite{Allada:2013nsw}, but
their $P_T$ values are too small (less than 0.7 GeV) to fix a large
scale.

We consider here the results of the HERMES Collaboration, selecting those
which best fulfil the kinematical conditions necessary for the validity of
our scheme, and compare them with our calculations based on TMD
factorisation and the Sivers and Collins functions extracted from SIDIS
data. In Section~\ref{For} we briefly summarise our formalism and in
Section~\ref{Mod} we compare our numerical results with data and give some
predictions for future measurements. Some final comments are given in
Section~\ref{comm}.

\section{\label{For} Formalism}

In Ref.~\cite{Anselmino:2009pn} (to which we refer for all details) we
considered the process $\pup \ell \to h \, X$ in the proton-lepton {\it c.m.}
frame (with the polarised proton moving along the positive $Z_{cm}$ axis)
and the transverse Single Spin Asymmetry:
\be
A_N = \frac{d\sigma^\uparrow(\bfP_T) - d\sigma^\downarrow(\bfP_T)}
           {d\sigma^\uparrow(\bfP_T) + d\sigma^\downarrow(\bfP_T)}
    = \frac{d\sigma^\uparrow(\bfP_T) - d\sigma^\uparrow(-\bfP_T)}
           {2 \, d\sigma^{\rm unp}(\bfP_T)} \,, \label{an}
\ee
where
\be
d\sigma^{\uparrow, \downarrow} \equiv \frac{E_h \, d\sigma^{p^{\uparrow,
\downarrow} \, \ell \to h\, X}}{d^{3} \bfP_h}
\ee
is the cross section for the inclusive process $p^{\uparrow, \downarrow}
\, \ell \to h \, X$ with a transversely polarised proton with spin ``up"
($\uparrow$) or ``down" ($\downarrow$) with respect to the scattering
plane~\cite{Anselmino:2009pn}.
$A_N$ can be measured either by looking at the production of hadrons
at a fixed transverse momentum $\bfP_T$, changing the incoming proton
polarisation from $\uparrow$ to $\downarrow$, or keeping a fixed
proton polarisation and looking at the hadron production to the left
and the right of the $Z_{cm}$ axis (see Fig.~1 of Ref.~\cite{Anselmino:2009pn}). $A_N$ was defined (and computed)
for a proton in a pure spin state with a pseudo-vector polarisation
$\bfS_T$ normal ($N$) to the production plane and $|\bfS_T| = S_T = 1$.
For a generic transverse polarisation along an azimuthal direction $\phi_S$
in the chosen reference frame, in which the $\uparrow$ direction is given by
$\phi_S = \pi/2$, and a polarisation $S_T \not= 1$, one has:
\be
A(\phi_S, S_T) = \bfS_T \cdot (\hat{\bfp} \times \hat{\bfP}_T) \, A_N =
S_T \sin\phi_S \, A_N \>, \label{phis}
\ee
where $\bfp$ is the proton momentum. Notice that if one follows the usual
definition adopted in SIDIS experiments, one simply has:
\be
A_{TU}^{\sin\phi_S} \equiv \frac{2}{S_T} \,
\frac{\int \, d\phi_S \> [d\sigma(\phi_S) - d\sigma(\phi_S + \pi)]\> \sin\phi_S}
     {\int \, d\phi_S \> [d\sigma(\phi_S) + d\sigma(\phi_S + \pi)]}
= A_N \>.
\label{ATU}
\ee

Assuming the validity of the TMD factorisation scheme for the process
$p \,\ell \to h \, X$ in which the only large scale detected is the transverse
momentum $P_T$ of the final hadron in the proton-lepton {\it c.m.} frame, the
main contribution to $A_N$ comes from the Sivers and Collins
effects, and one has~\cite{Anselmino:2009pn,D'Alesio:2004up,Anselmino:2005sh,D'Alesio:2007jt}:
\be
A_N =
\frac
{{\displaystyle \sum_{q,\{\lambda\}} \int \frac{dx \, dz}
{16\,\pi^2 x\,z^2 s}}\;
d^2 \bfk_{\perp} \, d^3 \bfp_{\perp}\,
\delta(\bfp_{\perp} \cdot \hat{\bfp}'_q) \, J(p_\perp)
\> \delta(\hat s + \hat t + \hat u)
\> [\Sigma(\uparrow) - \Sigma(\downarrow)]^{q \ell \to q \ell}}
{{\displaystyle \sum_{q,\{\lambda\}} \int \frac{dx \, dz}
{16\,\pi^2 x\,z^2 s}}\;
d^2 \bfk_{\perp} \, d^3 \bfp_{\perp}\,
\delta(\bfp_{\perp} \cdot \hat{\bfp}'_q) \, J(p_\perp)
\> \delta(\hat s + \hat t + \hat u)
\> [\Sigma(\uparrow) + \Sigma(\downarrow)]^{q \ell \to q \ell}} \>,
\label{anh}
\ee
with
\bea
\sum_{\{\lambda\}}\,[\Sigma(\uparrow) - \Sigma(\downarrow)]^{q
\ell \to q \ell} &=& \frac{1}{2} \, \Delta^N\! f_{q/\pup}
(x,k_{\perp}) \cos\phi \, \left[\,|{\hat M}_1^0|^2 + |{\hat
M}_2^0|^2 \right] \,
D_{h/q} (z, p_{\perp})  \nonumber \\
&+& h_{1q}(x,k_{\perp}) \, \hat M_1^0 \hat M_2^0 \, \Delta^N\!
D_{h/\qup} (z, p_{\perp}) \, \cos(\phi' + \phi_q^h) \label{ds1}
\eea
and
\be
\sum_{\{\lambda\}}\,[\Sigma(\uparrow) +
\Sigma(\downarrow)]^{q \ell \to q \ell} =
f_{q/p} (x,k_{\perp}) \,
\left[\,|{\hat M}_1^0|^2 + |{\hat M}_2^0|^2 \right] \,
D_{h/q} (z, p_{\perp}) \>. \label{ss1}
\ee

All functions and all kinematical and dynamical variables appearing in the above equations are exactly defined in Ref.~\cite{Anselmino:2009pn} and its Appendices and in Ref.~\cite{Anselmino:2005sh}. We simply recall here their
meaning and physical interpretation.

\begin{itemize}
\item
$\bkt$ is the transverse momentum of the parton in the proton and $\bpp$ is the transverse momentum of the final hadron with respect to the direction of the fragmenting parent parton, with momentum $\bfp^\prime_q$. $\phi$ is the
azimuthal angle of $\bkt$.
\item
The first term on the r.h.s.~of Eq.~(\ref{ds1}) shows the contribution to $A_N$ of the Sivers function $\Delta^N\!
f_{q/\pup}(x,k_\perp)$~\cite{Sivers:1989cc,Sivers:1990fh,Bacchetta:2004jz},
\bea
\Delta \hat f_{q/p,S}(x, \bfk_{\perp}) = \hat f_{q/p,S}(x,
\bfk_{\perp}) - \hat f_{q/p,-S}(x, \bfk_{\perp}) &\equiv& \Delta^N\!
f_{q/\pup}\,(x, k_{\perp}) \>
\bfS_T \cdot (\hat{\bfp} \times \hat{\bfk}_{\perp }) \label{defsivnoi} \\
&=& -2 \, \frac{k_\perp}{M} \, f_{1T}^{\perp q}(x, k_{\perp}) \>
\bfS_T \cdot (\hat{\bfp} \times \hat{\bfk}_{\perp }) \>. \nonumber
\eea
It couples to the unpolarised elementary interaction ($\propto (|M_1^0|^2 + |M_2^0|^2)$) and the unpolarised fragmentation function $D_{h/q} (z, p_{\perp})$; the $\cos\phi$ factor arises from the $\bfS_T \cdot (\hat{\bfp} \times \hat{\bfk}_{\perp })$ correlation factor.
\item
The second term on the r.h.s.~of Eq.~(\ref{ds1}) shows the contribution to $A_N$ of the unintegrated transversity distribution $h_{1q}(x,k_{\perp})$ coupled to the Collins function $\Delta^N\! D_{h/\qup} (z, p_{\perp})$~\cite{Collins:1992kk,Bacchetta:2004jz},
\bea
\Delta \hat D_{h/q^\uparrow}\,(z, \bfp_{\perp}) = \hat
D_{h/q^\uparrow}\,(z, \bfp_{\perp}) - \hat D_{h/q^\downarrow}\,(z,
\bfp_{\perp}) &\equiv& \Delta^N\! D_{h/\qup}\,(z, p_{\perp}) \>
\bfs_q \cdot (\hat{\bfp}_q^\prime \times \hat{\bfp}_{\perp }) \\
\label{defcolnoi} &=& \frac{2 \, p_\perp}{z \, m_h} H_{1}^{\perp q}(z,
p_{\perp}) \> \bfs_q \cdot (\hat{\bfp}_q^\prime \times
\hat{\bfp}_{\perp}) \>. \nonumber
\eea
This non perturbative effect couples to the spin transfer elementary interaction
($d\hat\sigma^{q^\uparrow \ell \to q^\uparrow \ell} -
d\hat\sigma^{q^\uparrow \ell \to q^\downarrow \ell}
\propto \hat M_1^0 \, \hat M_2^0$). The factor $\cos(\phi' + \phi_q^h)$ arises
from phases in the $\bfk_\perp$-dependent transversity distribution, the Collins
function and the elementary polarised interaction.
\end{itemize}

Some final comments on the kinematical configuration and the notations adopted in the HERMES experiment, with respect to those of Ref.~\cite{Anselmino:2009pn}, are necessary. According to the usual conventions adopted for SIDIS processes, in HERMES paper~\cite{Airapetian:2013bim} the lepton is assumed to move along the positive $Z_{cm}$ axis, so that the processes we are considering here are $\ell \, \pup \to h \, X$, rather than $\pup \ell \to h \, X$. In this reference frame the $\uparrow$ ($\downarrow$) direction is still along the $+Y_{cm}$ ($-Y_{cm}$) axis as in Ref.~\cite{Anselmino:2009pn} and, keeping the usual definition of $x_F = 2 P_L/\sqrt s$, where $P_L$ is the longitudinal momentum of the final hadron, only its sign is reversed.

The azimuthal dependent cross section measured by HERMES is defined
as~\cite{Airapetian:2013bim}:
\be
d\sigma = d\sigma_{UU}[1+S_T \, A_{UT}^{\sin\psi} \sin\psi] \>,
\label{sigH}
\ee
where
\be
\sin \psi = \bfS_T \cdot (\hat{\bfP}_T \times \hat{\bfk})
\ee
coincides with our $\sin\phi_S$ of Eq.~(\ref{phis}), as $\bfp$ and $\bfk$
(respectively, the proton and the lepton 3-momenta) are opposite vectors
in the lepton-proton {\it c.m.} frame. Indeed, taking into account
that ``left" and ``right" are interchanged in Refs.~\cite{Anselmino:2009pn}
and \cite{Airapetian:2013bim} (as these are defined looking downstream along
opposite directions, respectively the proton and the lepton momentum
directions) and the definition of $x_F$, one has:
\be
A_{UT}^{\sin\psi}(x_F, P_T) = A_N^{\pup \ell \to h X}(-x_F, P_T)  \>,
\label{AUT-hermes}
\ee
where $A_N^{\pup \ell \to h X}$
is the SSA as given by Eq.~(\ref{anh}) and computed in Ref.~\cite{Anselmino:2009pn}, and $A_{UT}^{\sin\psi}$ is the quantity measured by HERMES~\cite{Airapetian:2013bim}.

\section{\label{Mod} Estimates for $A_{UT}^{\sin\psi}$, comparisons with data and predictions}

In this Section we present our estimates for $A_{UT}^{\sin\psi}$, following the notation and convention adopted by the HERMES experiment. In our computation,
based on the TMD factorisation, we consider two different sets of Sivers and
Collins functions (the latter coupled to the transversity distribution), as
previously obtained in a series of papers from fits of SIDIS and $e^+e^-$
data~\cite{Anselmino:2005ea,Anselmino:2007fs,Anselmino:2008sga,Anselmino:2008jk}.

These sets, besides some different initial assumptions, differ in the choice
of the collinear fragmentation functions (FFs). More precisely, we adopt the
Sivers functions extracted in Ref.~\cite{Anselmino:2005ea}, where
only up and down quark contributions were considered, together with the first
extraction of the transversity and Collins functions obtained in
Ref.~\cite{Anselmino:2007fs}. In such studies we adopted, and keep using here,
the Kretzer set for the collinear FFs~\cite{Kretzer:2000yf}. We shall refer to
this set of functions as the SIDIS~1 set.

We then consider a more recent extraction of the Sivers
functions~\cite{Anselmino:2008sga}, where also the sea quark contributions
were included, together with an updated version of the transversity and
Collins functions~\cite{Anselmino:2008jk}; in these cases we adopted another
set for the FFs, namely that one by de Florian, Sassot and Stratmann (DSS)~\cite{deFlorian:2007aj}. We shall refer to
these as the SIDIS~2 set.

The use of these two sets of parameterisations, with their peculiar differences,
allows to take into account both the role of different weights between leading
and non-leading collinear FFs, as well as the different behaviour in the large
$x$ region of the Sivers and transversity distributions.
The large $x$ behaviour of these functions is still largely unconstrained by
SIDIS data, while it might be relevant to explain the values of $A_N$ measured
in $\pup p \to \pi\,X$ processes at RHIC, as studied in
Refs.~\cite{Anselmino:2013rya,Anselmino:2012rq}.
As this paper focuses on a process kinematically much closer to SIDIS, the
large $x$ behaviour of the involved TMDs is not so relevant here. Our two sets
of TMDs (SIDIS 1 and SIDIS 2) are well representative of the possible
uncertainties.

We then simply compute the values of $A_N$ as resulting, in the TMD factorised
scheme, from the -- {\it SIDIS and $e^+e^-$ extracted} -- SIDIS 1 and SIDIS 2
sets of TMDs. We will also show the uncertainty bands obtained by combining the statistical uncertainty bands of the Sivers and Collins functions, given
by the procedure described in Appendix A of Ref.~\cite{Anselmino:2008sga}.

In the following we will consider both the fully inclusive data from
$\ell \, p \to \pi \, X$ processes at large $P_T$, as well as the sub-sample
of data from processes in which also the final lepton is tagged (SIDIS category).
In the first case there is only one large scale, the $P_T$ of the final pion,
and for $P_T \simeq 1$ GeV, in order to avoid the low $Q^2$ region, one has to
look at pion production in the backward proton hemisphere (according to the
HERMES conventions this means $x_F > 0$). In this region (large $P_T$ and
$x_F > 0$) the lepton-quark scattering is still dominated by $Q^2 \ge 1$
GeV$^2$ and our pQCD computation is under control.

For the tagged-lepton sub-sample data $Q^2$ is measured and chosen to be
always bigger than 1 GeV$^2$. Notice that even in such conditions, working
in the lepton-proton {\it c.m.} frame, $P_T$ is still defined as the transverse
momentum of the pion w.r.t.~the lepton-proton direction. We will refer to
these data as ``SIDIS category".

Another important aspect to keep in mind is that in both cases (inclusive or
lepton-tagged events) one is not able to separate the single contributions
to $A_N$ of the Sivers and Collins effects, that in principle could contribute
together.

\subsection{Fully inclusive case}

\begin{figure}[h!t]
\includegraphics[width=6.truecm,angle=0]{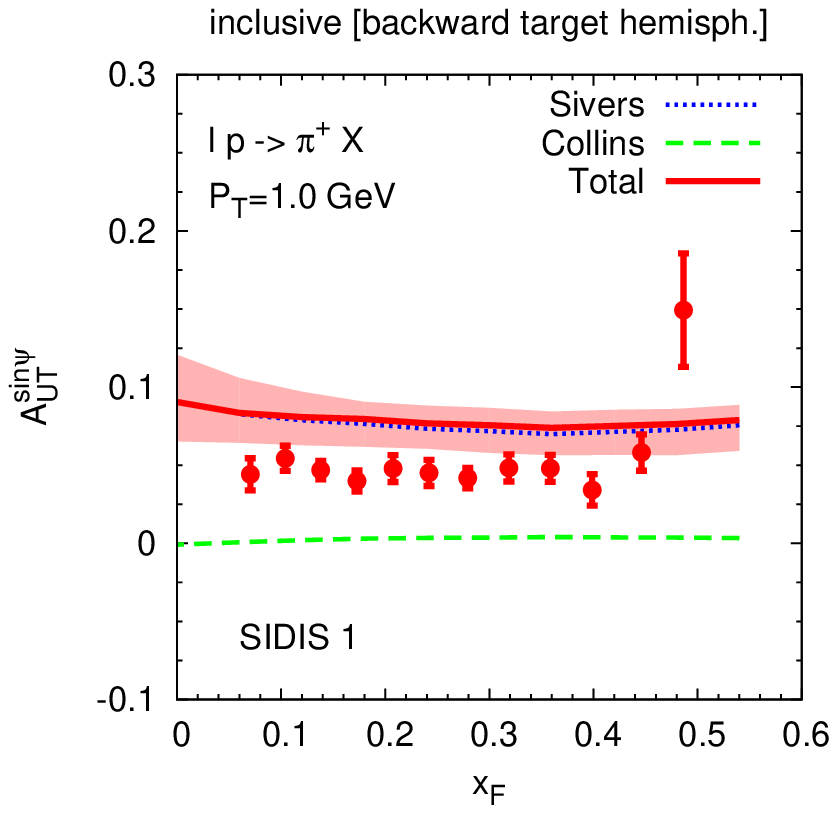}
\includegraphics[width=6.truecm,angle=0]{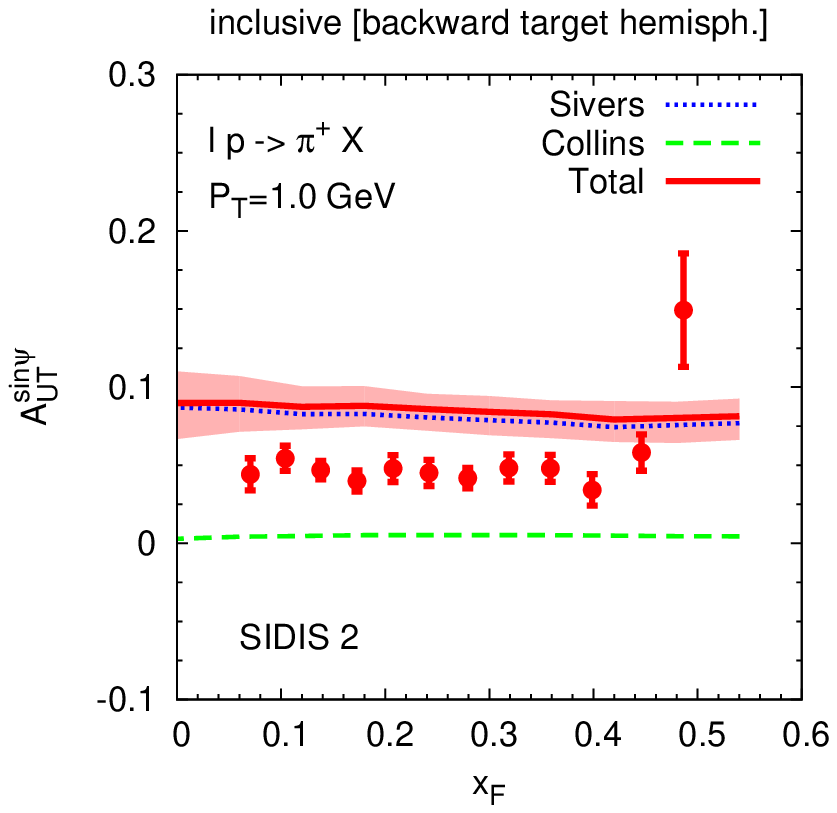}
\caption{The theoretical estimates for $A_{UT}^{\sin\psi}$ vs.~$x_F$ at
$\sqrt{s}\simeq 7$ GeV and $P_T = 1$ GeV for inclusive $\pi^+$ production in
$\ell \, \pup \to \pi \, X$ processes, computed according to
Eqs.~(\ref{AUT-hermes}) and (\ref{anh})--(\ref{ss1}) of the text, are compared
with the HERMES data~\cite{Airapetian:2013bim}. The contributions from the
Sivers (dotted blue lines) and the Collins (dashed green lines) effects are
shown separately and also added together (solid red lines). The computation
is performed adopting the Sivers and Collins functions of
Refs.~\cite{Anselmino:2005ea, Anselmino:2007fs}, referred to as SIDIS~1 in
the text (left panel), and of Refs.~\cite{Anselmino:2008sga, Anselmino:2008jk},
SIDIS~2 in the text (right panel). The overall statistical uncertainty band,
also shown, is the envelope of the two independent statistical uncertainty
bands obtained following the procedure described in Appendix A of
Ref.~\cite{Anselmino:2008sga}.}
\label{fig:an-hermes-pip}
\end{figure}

\begin{figure}[h!t]
\includegraphics[width=6.truecm,angle=0]{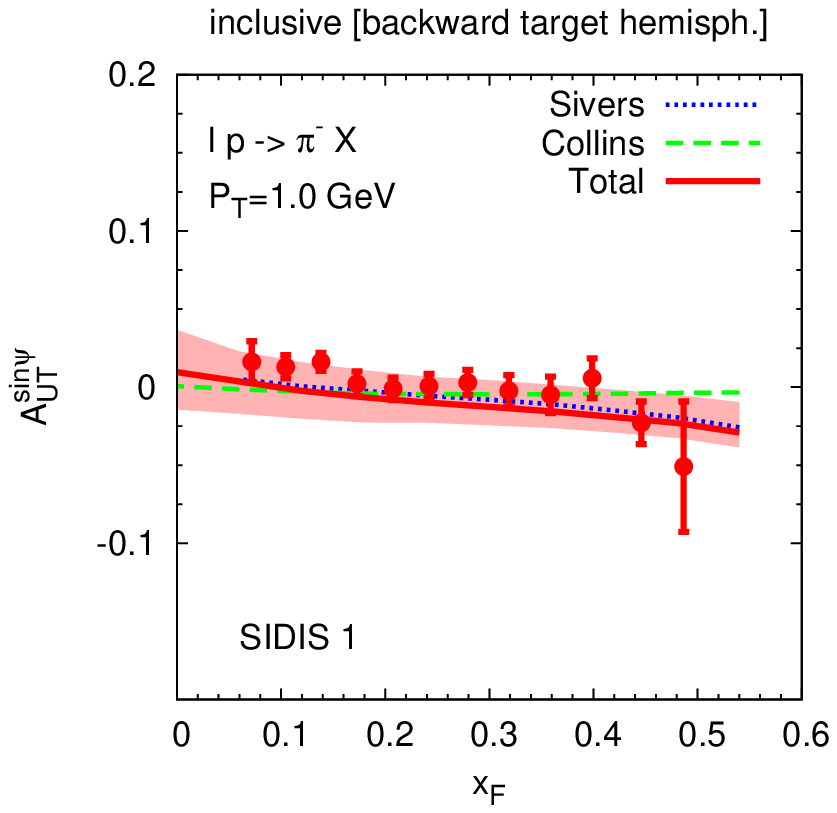}
\includegraphics[width=6.truecm,angle=0]{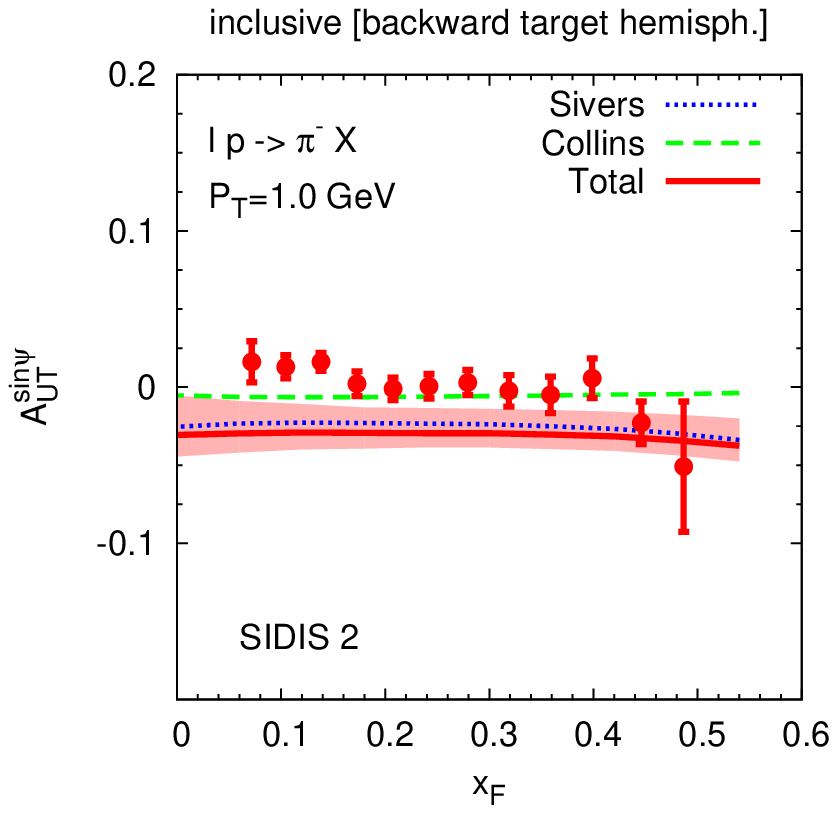}
\caption{Same as in Figure~\ref{fig:an-hermes-pip} but for inclusive $\pi^-$ production.}
\label{fig:an-hermes-pim}
\end{figure}

In this case, in order to apply our TMD factorised approach, one has to
consider data at large $P_T$. Among the HERMES data there is one bin that
fulfils this requirement, with $1 \lsim P_T \lsim 2.2$ GeV, and $\langle
P_T \rangle \simeq$ 1--1.1 GeV. In Figs.~\ref{fig:an-hermes-pip} and
\ref{fig:an-hermes-pim} we show a comparison of our estimates with these data,
respectively for positive and negative pion production. More precisely, we show
the results coming from both sets of TMDs, SIDIS 1 (left panels) and SIDIS 2
(right panels), for the Sivers (dotted blue lines) and Collins (dashed green
lines) effects separately, together with their sum (solid red lines). We also
computed the statistical uncertainty bands for both effects and showed the
envelope of the two bands (shaded area). Some comments are in order:
\begin{itemize}
\item
In this kinematical region the Collins effect is always negligible,
almost compatible with zero. The reason is twofold: from one side the partonic
spin transfer in the backward proton hemisphere is dynamically suppressed,
as explained in Ref.~\cite{Anselmino:2009pn}; secondly, the azimuthal phase
(see the second term on the r.h.s.~of Eq.~(\ref{ds1})) oscillates strongly, washing out the
effect.
\item
The Sivers effect does not suffer from any dynamical suppression, since it
enters with the unpolarised partonic cross section. Moreover, there is no
suppression from the integration over the azimuthal phases, as it happens,
for instance, in $p \, p \to \pi \, X$ case. Indeed in $\ell \, p \to \pi \, X$
only one partonic channel is at work and, for the moderate $Q^2$ values
of HERMES kinematics, the Sivers phase ($\phi$) appearing in the first
term on the r.h.s.~of Eq.~(\ref{ds1}) appears also significantly in the elementary
interaction, thus resulting in a non-zero phase integration.
\item
Moreover, in this kinematical region, even if looking at the backward hemisphere
of the polarised proton, one probes its valence region, where the extracted
Sivers function are well constrained. The reason is basically related to the
moderate {\it c.m.} energy, $\sqrt s \simeq 7$ GeV, of the HERMES experiment.
\item
The difference between SIDIS~1 and SIDIS~2 results for the negative
pion case, Fig.~\ref{fig:an-hermes-pim}, comes from the fact that in the
first case the Sivers function for up quark plays a relative bigger role,
even if coupled with the non-leading FF.
\item
The results presented here for the SIDIS 2 set of TMDs correspond to the
predictions given in Ref.~\cite{Anselmino:2009pn}, with the difference that
they were obtained for $P_T = 1.5$ and 2.5 GeV, and one should change $x_F$
into $-x_F$.
\end{itemize}

As one can see, while the SSA for positive pion production is a bit overestimated,
Fig.~\ref{fig:an-hermes-pip}, the description of the negative pion SSAs is in fair
agreement with data for the SIDIS 1 set (left panel in Fig.~\ref{fig:an-hermes-pim}).
Notice that in the fully inclusive case under study, at such values of $\sqrt s$
and $Q^2$ other effects could contaminate the SSA. Nonetheless the qualitative description, in size, shape and sign, is quite encouraging.


\subsection{Tagged or semi-inclusive category}

The HERMES Collaboration presents also a sub-sample of $\ell \, p$ data where
the final lepton is tagged~\cite{Airapetian:2013bim}. Of course the number of
these events is strongly reduced w.r.t.~the fully inclusive case. Nonetheless
the observed asymmetries are sizeable and show a peculiar behaviour.

We then consider also these data by imposing HERMES cuts: $Q^2>1$ GeV$^2$,
$W^2>10$ GeV$^2$, $0.023< x_B< 0.4$, $0.1<y<0.95$ and $0.2<z_h<0.7$, where
these are the standard variables adopted for the study of SIDIS processes.
Even in this case we restrict the analysis to the large $P_T$ region, namely
$P_T > 1$ GeV. In fact, in contrast to the SIDIS azimuthal asymmetries analysed
in the $\gamma^*-p\,$ {\it c.m.} frame, where the low $P_T\le 1$ GeV of the final
hadron is entirely given at leading order in terms of the intrinsic transverse
momenta in the distribution and fragmentation functions, here, working in the
$\ell-p\,$ {\it c.m.} frame, the observed $P_T$ is also given by the hard
scattering process. For this reason, to be sensitive to the intrinsic transverse
momentum effects, one has not to consider necessarily very small $P_T$ values.

\begin{figure}[h!t]
\includegraphics[width=6.truecm,angle=0]{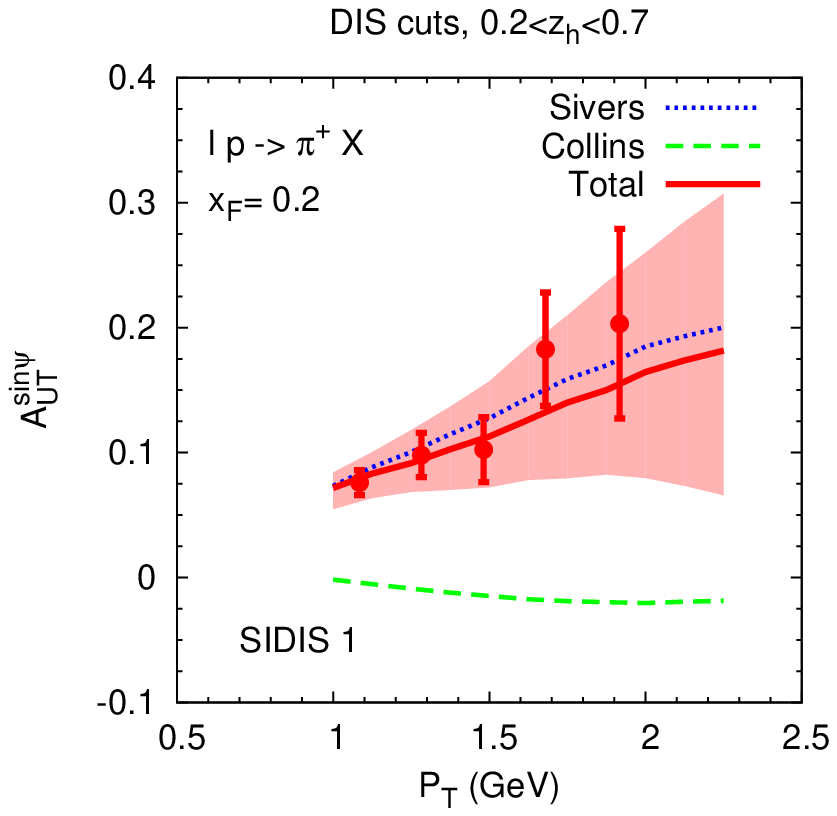}
\includegraphics[width=6.truecm,angle=0]{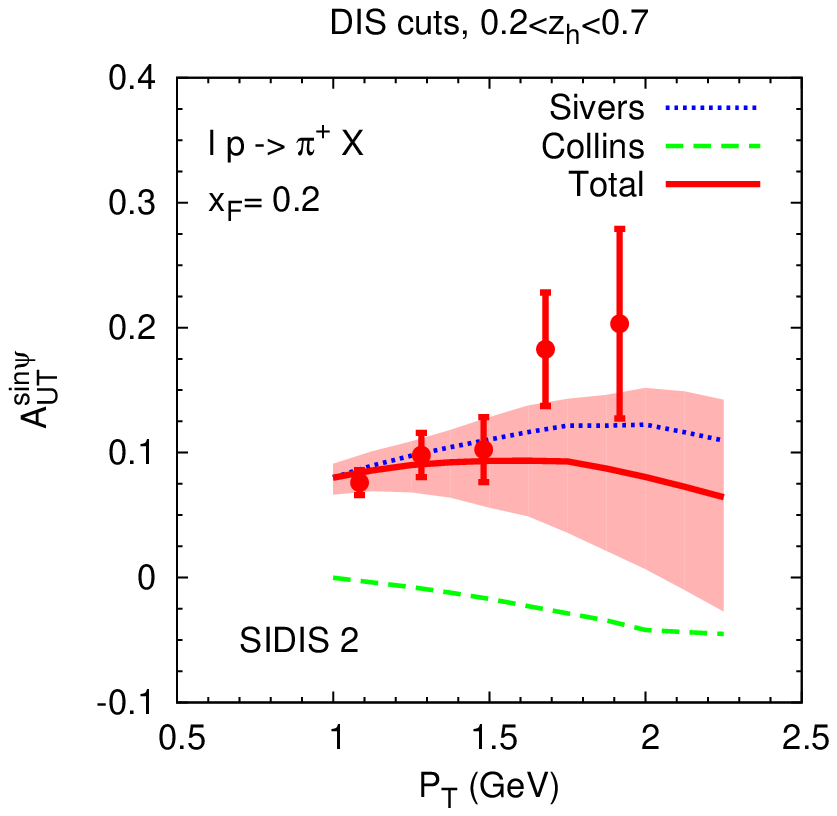}
\caption{
The theoretical estimates for $A_{UT}^{\sin\psi}$ vs.~$P_T$ at $\sqrt{s}\simeq
7$ GeV and $x_F = 0.2$ for inclusive $\pi^+$ production for the lepton tagged
events in $\ell \, \pup \to \pi \, X$ process, computed according to
Eqs.~(\ref{AUT-hermes}) and (\ref{anh})--(\ref{ss1}) of the text, are compared
with the HERMES data~\cite{Airapetian:2013bim}.
The contributions from the Sivers (dotted blue lines) and the Collins (dashed
green lines) effects are shown separately and also added together (solid red
lines). The computation is performed adopting the Sivers and Collins functions
of Refs.~\cite{Anselmino:2005ea, Anselmino:2007fs}, referred as SIDIS~1 in the text
(left panel), and of Refs.~\cite{Anselmino:2008sga, Anselmino:2008jk}, SIDIS~2
in the text (right panel). The overall statistical uncertainty band, also shown,
is the envelope of the two independent statistical uncertainty bands obtained
following the procedure described in Appendix A of Ref.~\cite{Anselmino:2008sga}.
\label{fig:an-hermes-dis-pip}}
\end{figure}
\begin{figure}[h!t]
\includegraphics[width=6.truecm,angle=0]{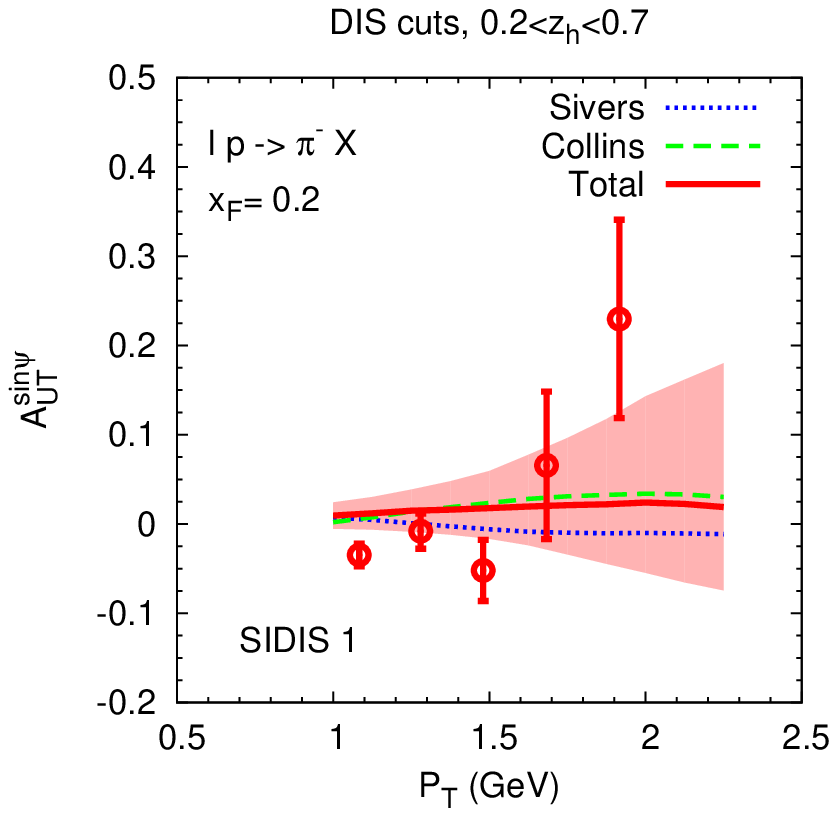}
\includegraphics[width=6.truecm,angle=0]{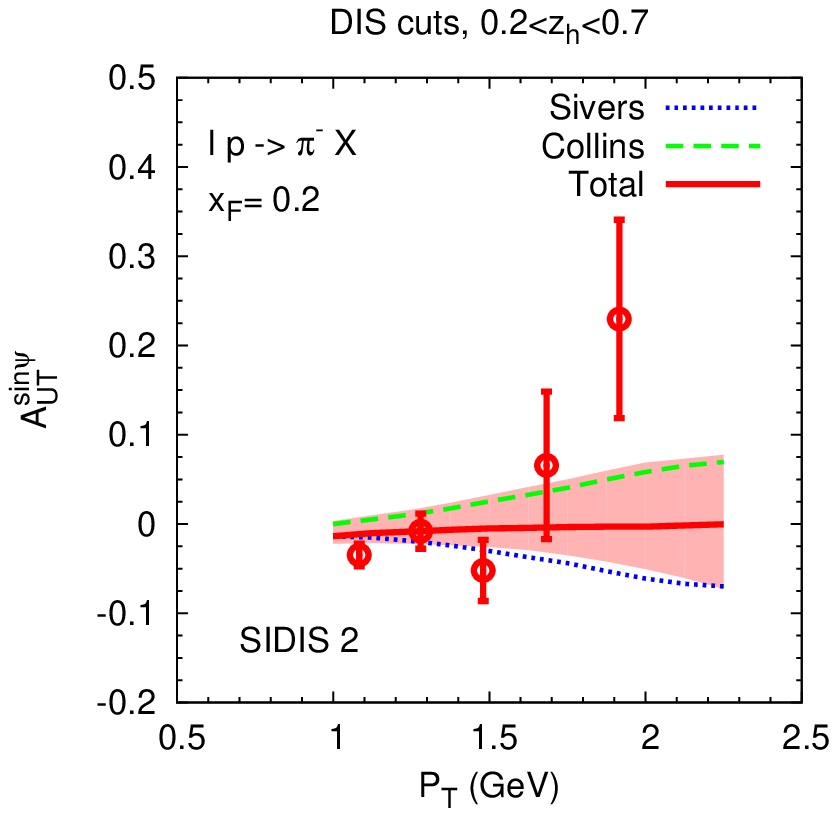}
\caption{Same as in Figure~\ref{fig:an-hermes-dis-pip} but for $\pi^-$ production.
\label{fig:an-hermes-dis-pim}}
\end{figure}

We show our estimates compared with HERMES data in
Figs.~\ref{fig:an-hermes-dis-pip} and \ref{fig:an-hermes-dis-pim},
respectively for positive and negative pion production as a function of $P_T$
at fixed $x_F = 0.2$. Notice that for $P_T > 1$ GeV, the values of $x_F$
probed in the HERMES kinematics are all very close to 0.2. We checked that
increasing the value of $x_F$, up to 0.3, the results are almost unchanged.
Again, we show the contributions from the Sivers (dotted blue line) and Collins
(dashed green line) effects separately and added together (solid red line) with
the overall uncertainty bands (shaded area). Some comments follow:
\begin{itemize}
\item
In this region the Collins effect (dashed green lines) is only partially suppressed by the dynamics and the azimuthal phase integration. Indeed the spin
transfer is still sizeable and the azimuthal phase entering the Collins effect
is peaked around $\pi$, that is the $\cos(\phi'+\phi_q^h)$ in the second term on the r.h.s.~of Eq.~(\ref{ds1}) is peaked around $-1$. Keeping in mind that the partonic spin
transfer is always positive and that the convolution of the transversity
distributions with the Collins functions is positive for $\pi^+$ and negative
for $\pi^-$, one can understand the sign of this contribution. The difference
between the SIDIS 1 and the SIDIS 2 sets (a factor around 2-3) comes from the
different behaviour of the quark transversity functions at moderately large $x$.
\item
The Sivers effect (dotted blue lines) for $\pi^+$ production
(Fig.~\ref{fig:an-hermes-dis-pip}) is sizeable for both sets. On the other
hand for $\pi^-$ production (Fig.~\ref{fig:an-hermes-dis-pim}) the SIDIS 1 set
(left panel) gives almost zero due to the strong cancellation between the
unsuppressed Sivers up quark distribution coupled to the non-leading FF, with
the more suppressed down quark distribution. In the SIDIS 2 set (right panel),
the same large $x$ behaviour of the up and down quark Sivers distributions
implies no cancellation.
\item
With the exception of the largest $P_T$ data point the description of the data
in terms of the sum of these effects is fairly good for both sets.
\end{itemize}

\subsection{Predictions}
\begin{figure}[h!t]
\includegraphics[width=6.truecm,angle=0]{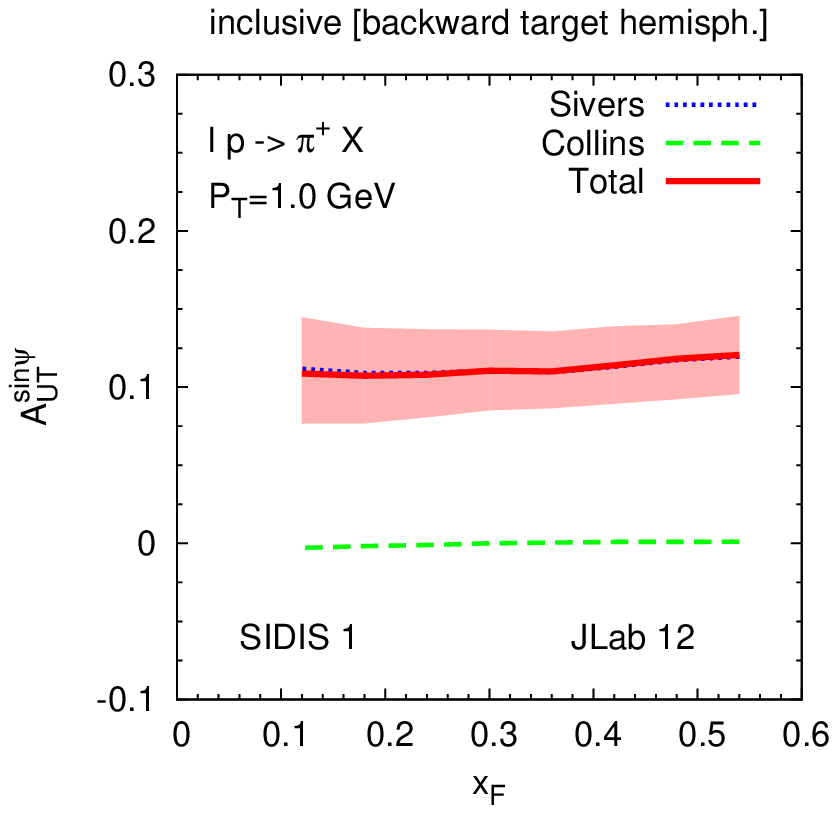}
\includegraphics[width=6.truecm,angle=0]{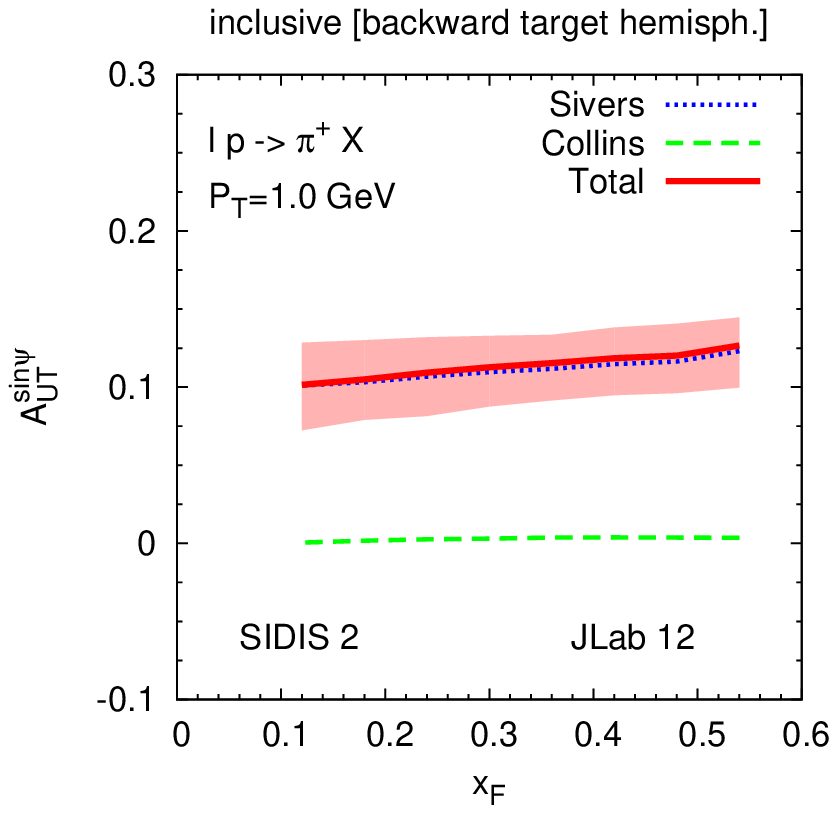}
\caption{Theoretical predictions for $A_{UT}^{\sin\psi}$ vs.~$x_F$ at
$\sqrt{s}\simeq 4.8$ GeV and $P_T = 1$ GeV for inclusive $\pi^+$ production in
$\ell \, \pup \to \pi \, X$ processes, computed according to
Eqs.~(\ref{AUT-hermes}) and (\ref{anh})--(\ref{ss1}) of the text, are shown
for future JLab experiments. The contributions from the
Sivers (dotted blue lines) and the Collins (dashed green lines) effects are
shown separately and also added together (solid red lines). The computation
is performed adopting the Sivers and Collins functions of
Refs.~\cite{Anselmino:2005ea, Anselmino:2007fs}, referred to as SIDIS~1 in
the text (left panel), and of Refs.~\cite{Anselmino:2008sga, Anselmino:2008jk},
SIDIS~2 in the text (right panel). The overall statistical uncertainty band,
also shown, is the envelope of the two independent statistical uncertainty
bands obtained following the procedure described in Appendix A of
Ref.~\cite{Anselmino:2008sga}.}
\label{fig:an-jlab-pip}
\end{figure}

\begin{figure}[h!t]
\includegraphics[width=6.truecm,angle=0]{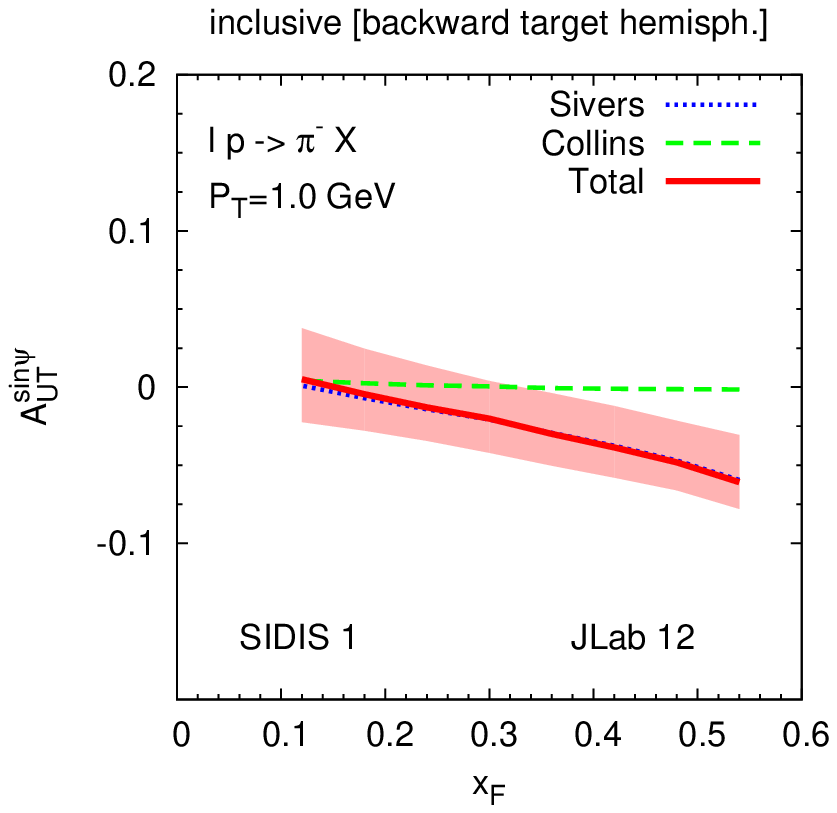}
\includegraphics[width=6.truecm,angle=0]{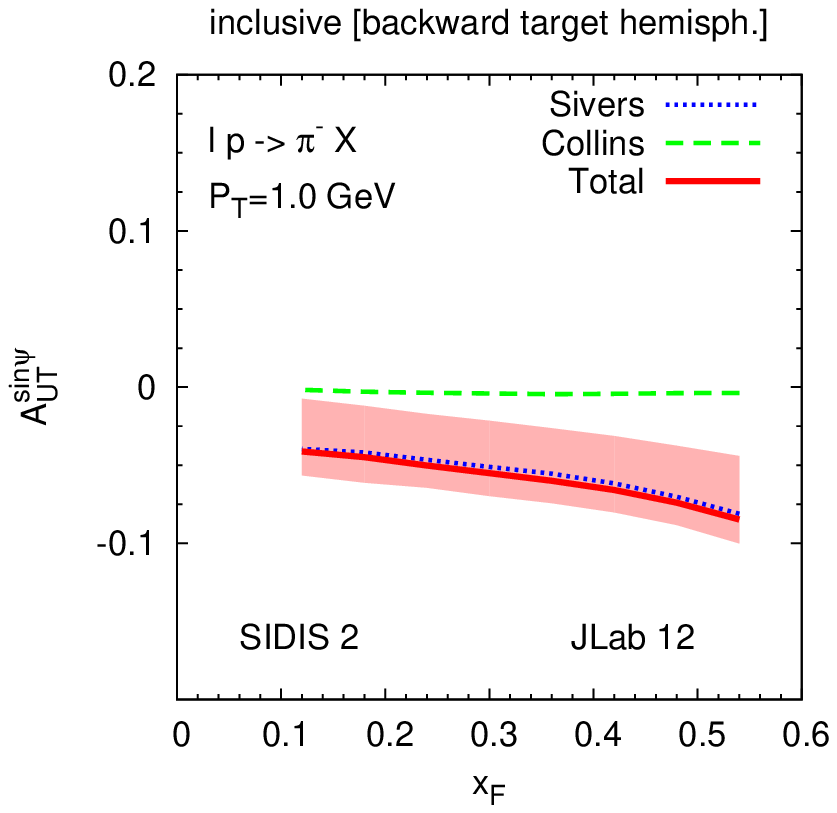}
\caption{Same as in Figure~\ref{fig:an-jlab-pip} but for inclusive $\pi^-$ production.}
\label{fig:an-jlab-pim}
\end{figure}

Data at $P_T \simeq 1$ GeV are expected from the future JLab 12 operation at
11 GeV. Because of the rather low  {\it c.m.} energy ($\sqrt s \simeq 4.8$ GeV),
in order to select data with large values of $Q^2$ one has to consider a
backward (w.r.t. the proton direction) production, which means $x_F \geq 0.1$.
With these kinematical bounds most contribution come from the quark valence
region. Our predictions, analogous to the results presented in
Figs.~\ref{fig:an-hermes-pip} and~\ref {fig:an-hermes-pim}, are shown in
Figs.~\ref{fig:an-jlab-pip} and~\ref {fig:an-jlab-pim}. The results
expected at JLab 12 are similar to those observed at HERMES.

\begin{figure}[h!t]
\includegraphics[width=6.truecm,angle=0]{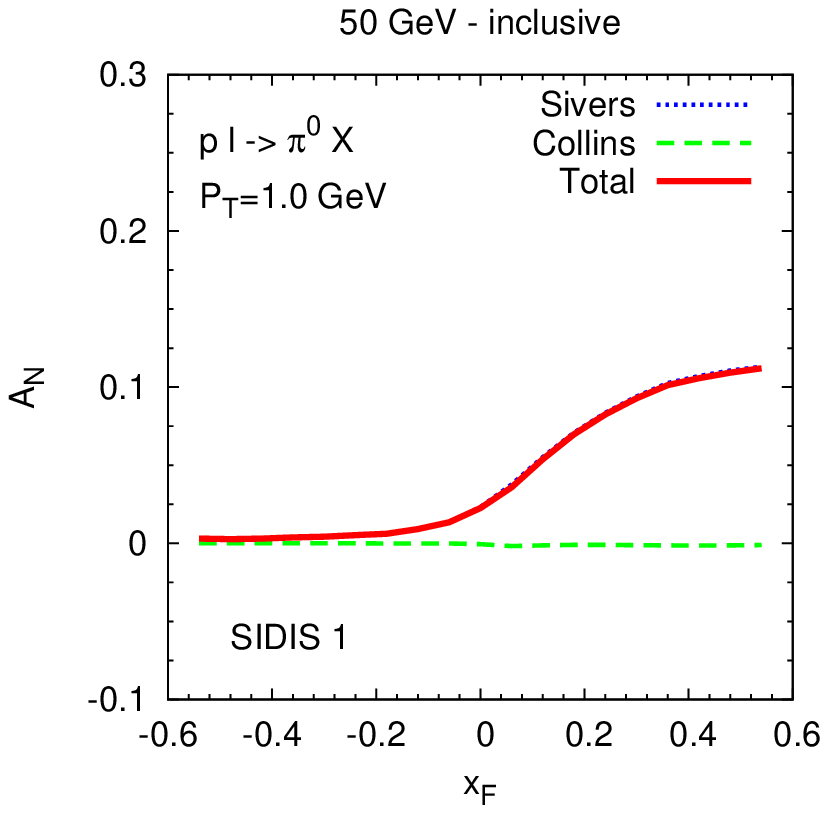}
\includegraphics[width=6.truecm,angle=0]{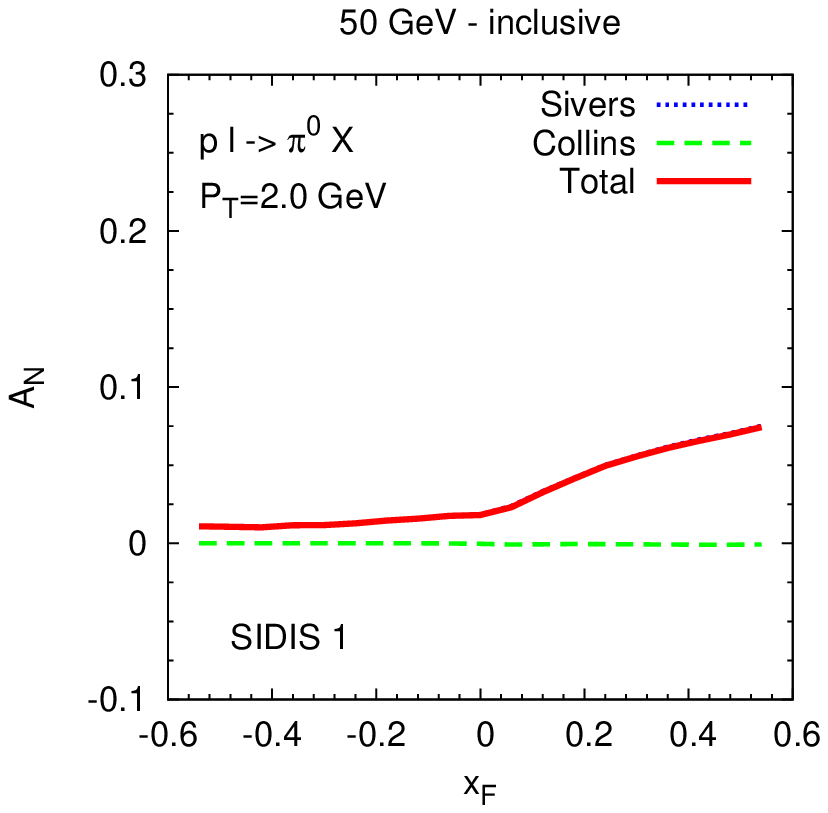}
\caption{
Theoretical estimates of $A_{N}$ vs.~$x_F$ at $\sqrt{s}\simeq 50$ GeV,
$P_T=1$ GeV (left panel) and $P_T=2$ GeV (right panel) for inclusive $\pi^0$
production in the $\pup \ell \to \pi \, X$ process. Notice that, contrary to the
kinematical configurations of Figs.~1 and 2, a forward production w.r.t.~the
proton direction corresponds here to positive values of $x_F$. The contributions
from the Sivers (dotted blue lines) and the Collins (dashed green lines) effects
are shown separately and also added together (solid red lines). The estimates are obtained adopting the Sivers and Collins functions of Refs.~\cite{Anselmino:2005ea, Anselmino:2007fs} (SIDIS~1 set), according to
Eqs.~(\ref{anh})--(\ref{ss1}) of the text.
\label{fig:an-50gev-pi0}}
\end{figure}

One might also wonder whether some features that characterise the SSAs observed
in $p \, p \to \pi \, X$ processes and that can be reproduced within a TMD
factorisation scheme~\cite{Anselmino:2013rya}, could still be encountered in
$\ell \, p \to \pi \, X$ reactions. To answer this question we consider the
inclusive $\ell \, p$ process at $\sqrt s= 50$ GeV. In this case, in order
to have a more direct comparison with the $\pup \, p$ case, we calculate $A_N$
as defined in Eq.~(\ref{ds1}), with the polarised proton moving along $Z_{cm}$,
that is with positive $x_F$ in the forward hemisphere of the polarised proton.
In Fig.~\ref{fig:an-50gev-pi0} we show our estimates of $A_N$ for $\pi^0$
production in the process $\pup \ell \to \pi \, X$ at $\sqrt s= 50$ GeV and
$P_T= 1$ GeV (left panel) and $P_T=2$ GeV (right panel) adopting the SIDIS 1 set.
This set indeed is the one that better reproduces the behaviour of $A_N$ in
$\pup p\to\pi\,X$ processes (see for instance Ref.~\cite{Boglione:2007dm}).
The result deserves a few comments.
\begin{itemize}
\item
The Collins effect in the backward region is totally negligible: this is due
to a strong suppression coming from the azimuthal phase integration. In the
forward region the SIDIS~1 set, as well the SIDIS~2 (results not shown), give
tiny values even if the azimuthal phase would be effective. In particular for
$x_F > 0.3$, once again the cosine factor entering this effect in
Eq.~(\ref{ds1}) is negative.
\item
The Sivers effect is sizeable and increasing with $x_F$ for positive values
of $x_F$, while negligible in the negative $x_F$ region. Notice that the
suppression of the Sivers effect for $x_F < 0$, even if in such a process
there is only one partonic channel, is due to a weak dependence on the
azimuthal phase of the elementary interaction at the large $Q^2$ values
reached at this energies.
\item
It is worth noticing that the functional shape of $A_N(x_F)$, for the
$\pup \ell \to h \, X$ large $P_T$ process, is similar to that
observed at various energies in $\pup p \to h \, X$ processes, being
negligible at negative $x_F$ and increasing with positive values of $x_F$.
\item
The process $\ell \, \pup \to {\rm jet} + X$ at large $\sqrt s$ values was
studied in a twist-3 formalism in Ref.~\cite{Kang:2011jw}. The quark-gluon-quark
Qiu-Sterman correlator $T_F$ was fixed exploiting its relation with the Sivers
function, taken from an extraction \cite{Anselmino:2008sga} from SIDIS data.
The value of $A_N$ was found to be positive for $x_F > 0$ (the same kinematical
configuration as for our Fig.~\ref{fig:an-50gev-pi0} was adopted), with results
very close to the results we find here for $\pi^0$ production and we found in
Ref.~\cite{Anselmino:2009pn} for jet production. Indeed the twist-3 and the
TMD mechanisms were shown to be closely related and provide a unified picture
for SSAs in SIDIS processes \cite{Ji:2006ub}. However, the factorised twist-3
collinear scheme, using the SIDIS extracted Sivers functions for fixing the
Qiu-Sterman correlator $T_F$, seems to have severe problems in explaining the
SSA $A_N$ observed in $p\,p$ processes, leading to values of $A_N$ opposite to
those measured~\cite{Kang:2011hk}. These issues were further studied in Refs.~\cite{Kang:2012xf,Metz:2012ui,Gamberg:2013kla}.
A recent analysis of $A_N$ in $p\,p$ scattering in the twist-3 formalism~\cite{Kanazawa:2014dca} attempts at solving this problem showing that the asymmetry might be dominated by new large effects
coming from fragmentation. It is not clear how much these same effects would
change the value of $A_N$ in SIDIS when going from jet to $\pi^0$ production.
\item
The measurement of asymmetries in the same kinematical region and with
the same features as in Fig.~\ref{fig:an-50gev-pi0} for $\pi^0$ production,
and as in Fig.~6 of Ref.~\cite{Anselmino:2009pn} for jet production, would be
a strong indication in support of our TMD factorised approach. Such measurements
might be possible at a future Electron-Ion-Collider (EIC)~\cite{Accardi:2012qut}.
\end{itemize}

\section{Comments and conclusions}\label{comm}

We have further pursued and tested the idea presented in
Ref.~\cite{Anselmino:2009pn} for assessing the validity of the TMD
factorisation in inclusive processes in which a single large $P_T$
particle is produced.
Starting from the TMD factorisation valid for SIDIS processes,
$\ell \, p \to \ell \, h \, X$, in which a large $Q^2$ virtual photon
$\gamma^*$ hits a quark, which then fragments into a final hadron
with a small $P_T$ in the $\gamma^*-p\,$ {\it c.m.} frame, we have assumed
its validity for processes in which the final lepton is not necessarily
observed, but the final detected hadron has a large $P_T$ in the
lepton-proton {\it c.m.} frame. A large value of $P_T$ implies, at leading order,
a large angle elementary scattering, $\ell \, q \to \ell \, q$, and then a
large value of $Q^2$. Such a process is analogous to the $\pup p \to h \, X$
processes, for which large SSAs $A_N$, Eq.~(\ref{an}), have been measured.
According to the TMD factorisation approach, the SSAs can be generated by
the Sivers and Collins effects~\cite{Anselmino:2013rya,Anselmino:2012rq}.

We have computed the single spin asymmetry $A_N$, for the $\ell \, \pup \to
h \, X$ process and in the TMD factorised scheme, as generated by the
Sivers and the Collins functions, which have been extracted from SIDIS and
$e^+e^-$ data~\cite{Anselmino:2005ea,Anselmino:2007fs,Anselmino:2008sga,
Anselmino:2008jk}. Doing so, we adopt a unified TMD factorised approach,
valid for $\ell \, p \to \ell \, h \, X$ and $\ell \, p \to h \, X$
processes, in which, consistently, we obtain information on the TMDs
and make predictions for $A_N$. Some of these predictions were given in
Ref.~\cite{Anselmino:2009pn}.

New HERMES data on $A_N$ are now available~\cite{Airapetian:2013bim} for
different kinematical regions; we have selected those data which -- although
not yet optimally -- fulfil the conditions of applicability of our
TMD factorisation approach, and compared them with the results of our
computations. We have selected two sets of TMDs extracted from SIDIS and
$e^+e^-$ data, and which are representative, with their large differences,
of the uncertainties which the SIDIS available data still allow.

It turns out, Figs.~\ref{fig:an-hermes-pip}--\ref{fig:an-hermes-dis-pim}, that
our theoretical estimates for $A_{UT}^{\sin\psi}(x_F, P_T) = A_N(-x_F, P_T)$
agree well, in shape and sign, with the experimental results, in particular for
one set of TMDs (SIDIS 1). In some cases (see Fig.~\ref{fig:an-hermes-pip}) our
results are a bit larger than data, yet with the right sign and behaviour; one
should not forget that in the kinematical regions we are considering (in $P_T$,
$Q^2$ and $\sqrt s$) other mechanisms might still be at work. The overall
agreement between our computations and the data is very encouraging.

In Figs.~\ref{fig:an-jlab-pip} and \ref{fig:an-jlab-pim} we have estimated
the expected value of $A_{UT}^{\sin\psi}$ at the future JLab experiments at
12 GeV. We are still in a kinematical region where a careful selection of
data is necessary in order to ensure the validity of our approach. The results
are similar to those obtained in Figs.~\ref{fig:an-hermes-pip}
and~\ref {fig:an-hermes-pim} for HERMES kinematics.

At last, in Fig.~\ref{fig:an-50gev-pi0}, we have given predictions for
$A_N(x_F)$ in very safe kinematical regions for our approach to hold. Indeed,
as expected, we recover the same behaviour for $A_N(x_F)$ as observed in
$\pup p \to \pi^0 \, X$ processes. Such a prediction, crucial for assessing the
validity of our TMD factorisation scheme, could be tested at a future
EIC~\cite{Accardi:2012qut}.

\acknowledgments
\noindent
A.P.~work is supported by the U.S.~Department of Energy under Contract No.~DE-AC05-06OR23177. \\
M.A., M.B., U.D., S.M.~and F.M.~acknowledge support from the European Community under the FP7 program ``Capacities - Research Infrastructures" (HadronPhysics3, Grant Agreement 283286). \\
M.A, M.B.~and S.M.~acknowledge support from the ``Progetto di Ricerca Ateneo/CSP" (codice TO-Call3-2012-0103). \\
U. D.~is grateful to the Department of Theoretical Physics II of the Universidad Complutense of Madrid for the kind hospitality extended to him during the completion of this work.



\end{document}